\documentclass[pss]{wiley2sp} 
\usepackage{amsmath}

\begin{document}

\title{%
Nonequilibrium dynamics of multiorbital correlated electron system
under time-dependent electric fields}

\titlerunning{%
Nonequilibrium dynamics of multiorbital correlated electron system}

\author{Hiroaki Onishi\textsuperscript{\Ast}}

\authorrunning{H. Onishi}

\mail{e-mail \textsf{onishi.hiroaki@jaea.go.jp}, Phone: +81-29-282-6729, Fax: +81-29-282-5939}

\institute{%
Advanced Science Research Center,
Japan Atomic Energy Agency, Tokai, Ibaraki 319-1195, Japan}

\received{XXXX, revised XXXX, accepted XXXX} 
\published{XXXX} 


\abstract{%
%
%
%
\abstcol{
To clarify a key role of orbital degrees of freedom
in the response of the many-body electron state of correlated electrons
to an external field,
we investigate the real-time dynamics
in an $e_{\rm g}$-orbital Hubbard model
under applied electric fields
by exploiting numerical techniques
such as a time-dependent density-matrix renormalization group (DMRG) method.
}
{
The ground state without applying an electric field is found to be
an antiferromagnetic/ferro-orbital state.
After we switch on an oscillating electric field,
it gives rise to the creation of holon-doublon pairs
when the frequency of the applied electric field exceeds a charge excitation gap.
We argue that a pair-hopping process yields
a dynamical deformation of the orbital configuration.
}
}

%
%

\maketitle   

\section{Introduction}

The photoinduced phase transition is a cooperative phenomenon
in nonequilibrium conditions driven by light illumination
\cite{Tokura2006}.
In general, in contrast to the conventional phase transition
at thermal equilibrium,
the system can get excited into a nonequilibrium state
that is not accessible via the control of external conditions
such as temperature, magnetic field, and pressure.
For a typical example in correlated electron systems,
Pr$_{0.7}$Ca$_{0.3}$MnO$_{3}$
shows a photoinduced insulator-to-metal transition
\cite{Miyano1997,Fiebig1998},
in which
a charge/orbital-ordered insulating state is melted
and a ferromagnetic metallic state is induced
in an ultrafast time scale.
Extensive theoretical efforts have been devoted to understand
the early stage dynamics and the relaxation dynamics
of the photoinduced phase transition processes
involving multiple degrees of freedom
\cite{Nasu2004,Takahashi2008,Koshibae2009,Onishi2010,Maeshima2010,Matsueda2012,Kanamori2012}.

In this paper, we study the nonequilibrium dynamics
of a multiorbital Hubbard model
under a time-dependent electric field,
by using numerical techniques.
We observe a dynamical charge excitation
accompanied by the creation of photocarriers.
We show that the pair-hopping process plays a significant role
in a dynamical deformation of the orbital configuration.

\section{Model and method}

Let us consider two $e_{\rm g}$ orbitals
in each site of a one-dimensional chain 
with $N$ sites along the $z$-axis.
The number of electrons per site is set to be one
(quarter filling).
The Hamiltonian of an orbital degenerate Hubbard model
including a time-dependent electric field is given by
\begin{eqnarray}
 H
 &=&
 \sum_{i,\tau,\tau',\sigma}
 -t_{\tau\tau'}
 ( {\rm e}^{{\rm i}A(t)} d_{i\tau\sigma}^{\dag} d_{i+1\tau'\sigma}
   +{\rm h.c.} )
 \nonumber\\
 &&
 +U \sum_{i,\tau} \rho_{i\tau\uparrow} \rho_{i\tau\downarrow}
 +U' \sum_{i,\sigma,\sigma'} \rho_{i\alpha\sigma} \rho_{i\beta\sigma'}
 \nonumber\\
 &&
 +J \sum_{i,\sigma,\sigma'}
 d_{i\alpha\sigma}^{\dag} d_{i\beta\sigma'}^{\dag}
 d_{i\alpha\sigma'} d_{i\beta\sigma}
 \nonumber \\
 &&
 +J' \sum_{i,\tau \ne \tau'} 
 d_{i\tau\uparrow}^{\dag} d_{i\tau\downarrow}^{\dag}
 d_{i\tau'\downarrow} d_{i\tau'\uparrow},
\label{eq-H}
\end{eqnarray}
where $d_{i\tau\sigma}$ is an annihilation operator
for an electron
with spin $\sigma$ (=$\uparrow$, $\downarrow$)
in orbital $\tau$
(=$\alpha$ for $3z^2$$-$$r^2$; $\beta$ for $x^2$$-$$y^2$)
at site $i$,
and
$\rho_{i\tau\sigma}$=$d_{i\tau\sigma}^{\dag}d_{i\tau\sigma}$.
$t_{\tau\tau'}$ is the electron hopping
between $\tau$ and $\tau'$ orbitals in nearest-neighbor sites,
given by
$t_{\alpha\alpha}$=$1$ (energy unit)
and $t_{\alpha\beta}$=$t_{\beta\alpha}$=$t_{\beta\beta}$=$0$
\cite{Slater1954}.
That is, the $3z^2$$-$$r^2$ orbital is itinerant,
while the $x^2$$-$$y^2$ orbital is localized.
Regarding the onsite interaction,
$U$, $U'$, $J$, and $J'$ are
the intraorbital Coulomb repulsion,
the interorbital Coulomb repulsion,
the interorbital exchange interaction,
and the pair-hopping interaction,
respectively.
Among four interaction parameters,
we assume that $U$=$U'$+$J$+$J'$ holds
due to the rotational invariance in the orbital space,
and $J$=$J'$ due to the reality of the wavefunction.
Thus we have two independent interaction parameters.
The electric field is described by
a time-dependent vector potential involved in the hopping term,
given by
\begin{equation}
 A(t) = \theta(t) A_{0} \sin \omega t,
\end{equation}
where $\theta(t)$ is the Heaviside step function,
i.e., we switch on the electric field at time $t$=0.
We use the unit such that $\hbar$=$1$,
and the time is measured in units of
$\hbar/t_{\alpha\alpha}$
\cite{Note1}.
In this paper,
we analyze the system with $N$=$16$ sites.
We fix $U'$=$10$ and study the behavior with varying $J'$ and $\omega$.
We mainly adopt $A_{0}$=$0.1$
and briefly discuss the dependence on $A_{0}$
in the weak perturbation regime.

We numerically investigate the real-time dynamics
of the model (\ref{eq-H})
by exploiting
time-dependent density-matrix renormalization group (DMRG) techniques
\cite{White1992,Daley2004,White2004,Schollwock2005a,Schollwock2005b}.
First, as an initial state at $t$=$0$,
we obtain the ground state without the electric field
by an ordinary static DMRG method with the use of
the finite-system algorithm under open boundary conditions.
Then,
after the electric field is switched on at $t$=$0$,
the time evolution of the wavefunction is computed
by an adaptive time-dependent DMRG method,
based on the second-order Suzuki-Trotter decomposition
with a small time step.
We keep up to $m$=$600$ states
and the truncation error is kept around $10^{-7}$
during the time evolution.

\section{Numerical results}

Let us first discuss the electron configuration
of the initial state before the electric field is switched on
(see Fig.~1).
Since electrons favorably occupy the itinerant orbital,
the $3z^2$$-$$r^2$ orbital is singly occupied at every site,
indicating a charge- and orbital-ordered state.
An antiferromagnetic exchange interaction among $S$=$1/2$ spins
yields an antiferromagnetic state.
Here, we apply the electric field
to cause a nonequilibrium charge excitation,
in which we expect the creation of photocarriers
called holons and doublons.
To gain an insight into the dynamical change of the charge state,
we measure the time evolution of a doublon number,
defined by
\begin{equation}
 N_{\rm d} = \sum_{i} \langle n_{i\uparrow}n_{i\downarrow} \rangle_{t},
\end{equation}
where $\langle\cdots\rangle_{t}$ denotes the expectation value
using the wavefunction at time $t$.
In Fig.~2(a), we present $N_{\rm d}$ at $J'$=$0$.
We find that for small $\omega$,
the doublon number shows just a small fluctuation
around the ground-state value,
since the energy of the electric field is not large enough to
cause the charge excitation across a charge gap,
defined by
\begin{equation}
 \Delta_{\rm c}(N,N_{\rm e}) =
 E(N,N_{\rm e}+1)+E(N,N_{\rm e}-1)-2E(N,N_{\rm e}),
\end{equation}
where $E(N,N_{\rm e})$ denotes the lowest energy of
the $N$-site system with $N_{\rm e}$ electrons
before applying the electric field.
In the present case, $N_{\rm e}$=$N$ at quarter filling.
The charge gap is estimated to be $6.70$
for $N$=$16$, $U'$=$10$, and $J'$=$0$.
As we increase $\omega$,
the doublon number increases with time
when $\omega$ exceeds a threshold value.
Note that the threshold is at around $\omega$$\sim$$7$,
which agrees with the charge gap.
Thus the electric field induces the dynamical charge excitation
accompanied by the creation of photo-carriers.

\begin{figure}[t]
\includegraphics*[width=\linewidth]{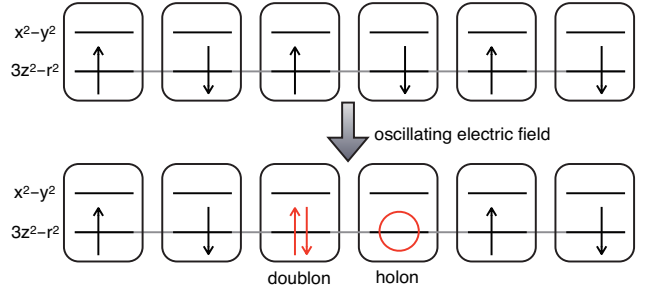}
\caption{%
The electron configuration in the ground state
and the charge excited state with a holon-doublon pair.
}
\label{fig1}
\end{figure}

To clarify the effects of the emergent dynamical charge excitation
on the spin and orbital states,
we measure spin and orbital structure factors, defined by
\begin{equation}
 S(q) =
 (1/N^2) \sum_{j,k}
 {\rm e}^{{\rm i}q(j-k)}
 \langle S_{j}^{z} S_{k}^{z} \rangle_{t},
\end{equation}
\begin{equation}
 T(q) =
 (1/N^2) \sum_{j,k}
 {\rm e}^{{\rm i}q(j-k)}
 \langle T_{j}^{z} T_{k}^{z} \rangle_{t},
\end{equation}
with
$S_i^z$=$\sum_{\tau}(\rho_{i\tau\uparrow}$$-$$\rho_{i\tau\downarrow})/2$
and
$T_i^z$=$\sum_{\sigma}(\rho_{i\alpha\sigma}$$-$$\rho_{i\beta\sigma})/2$.
Figures~2(b) and (c) show
$S(q)$ and $T(q)$, respectively, at $J'$=$0$.
As for the spin state,
$S(q)$ has an antiferromagnetic peak at $q$=$\pi$.
We observe that
$S(\pi)$ gradually decreases for large $\omega$,
while there is no change for small $\omega$,
indicating that
the antiferromagnetic correlation is suppressed
with the progress of the dynamical charge excitation.
On the other hand,
since the $3z^2$$-$$r^2$ orbital is singly occupied in the ground state,
$T(q)$ has a sharp ferro-orbital peak at $q$=$0$ at $t$=$0$.
It turns out that $T(q)$ does not change with time at all irrespective of $\omega$.
We can understand the origin of this characteristic behavior as follows.
After the electric field is switched on,
holon-doublon pairs are created through the electron hopping process.
Here, we should note again that
the $3z^2$$-$$r^2$ orbital is itinerant,
while the $x^2$$-$$y^2$ orbital is localized.
In such a situation, doublons are introduced into the $3z^2$$-$$r^2$ orbital,
as shown in Fig.~1.
During the time evolution,
the electron hopping process causes the doublon motion
only between $3z^2$$-$$r^2$ orbitals.
We also remark that the pair-hopping process would yield the doublon motion
between $3z^2$$-$$r^2$ and $x^2$$-$$y^2$ within a site,
but we have no process to change the orbital state at $J'$=$0$.
Thus, the ferro-orbital state is not affected
even when the dynamical charge excitation occurs.

\begin{figure}[t]
\includegraphics*[width=\linewidth]{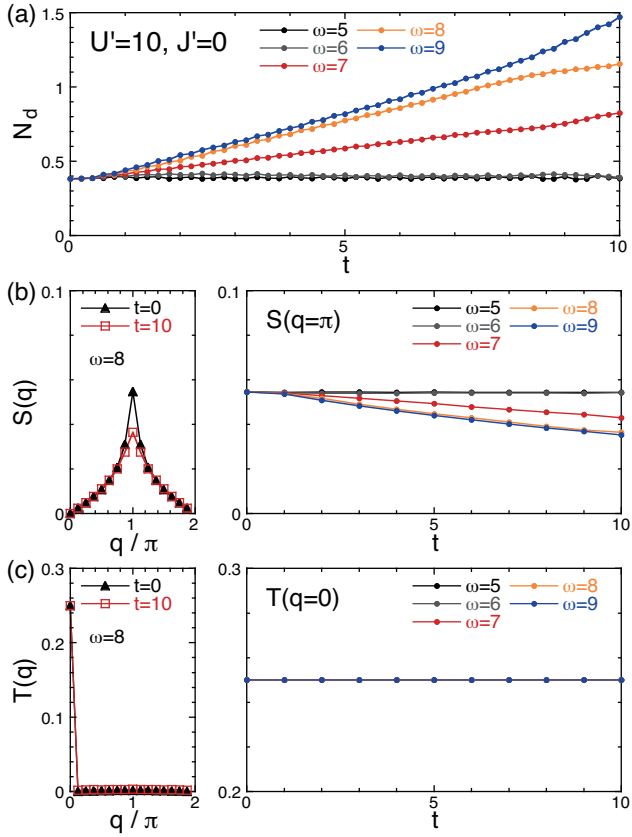}
\caption{%
The time evolution of
(a) the doublon number $N_{\rm d}$,
(b) the spin structure factor $S(q)$,
and
(c) the orbital structure factor $T(q)$
for $N$=$16$, $U'$=$10$, $J'$=$0$, and $A_{0}$=$0.1$.
}
\label{fig2}
\end{figure}

\begin{figure}[t]
\includegraphics*[width=\linewidth]{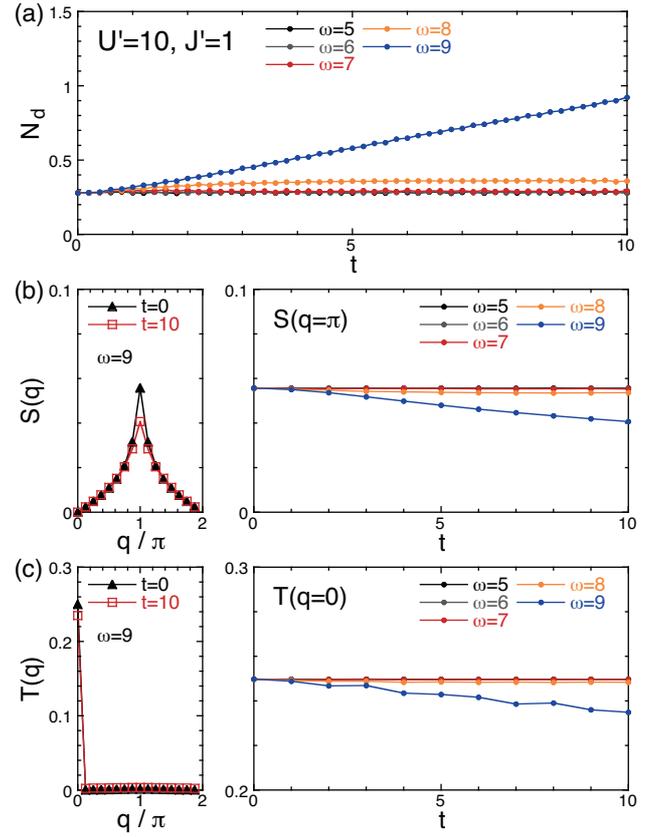}
\caption{%
The time evolution of
(a) the doublon number $N_{\rm d}$,
(b) the spin structure factor $S(q)$,
and
(c) the orbital structure factor $T(q)$
for $N$=$16$, $U'$=$10$, $J'$=$1$, and $A_{0}$=$0.1$.
}
\label{fig3}
\end{figure}

Now let us turn our attention to the case of finite $J'$.
In Fig.~3, we present $N_{\rm d}$, $S(q)$, and $T(q)$ at $J'$=$1$.
In Fig.~3(a), we observe that
$N_{\rm d}$ increases with time when $\omega$ exceeds a threshold,
in a similar way to the case of $J'$=$0$.
However, the threshold is at around $\omega$$\sim$$8$,
and it is somewhat larger than the charge gap estimated to be $6.43$,
implying that
the electric field causes
a high-energy excitation rather than the lowest-energy excitation.
Detailed analysis to reveal how the threshold and the charge gap
behave with varying $J'$ is an interesting future problem.
Regarding the spin state,
$S(q)$ also exhibits a similar behavior to the case of $J'$=$0$,
as shown in Fig.~3(b).
The antiferromagnetic correlation is gradually suppressed with time
according to the dynamical charge excitation.
On the other hand,
Fig.~3(c) shows a qualitatively different behavior
from the case of $J'$=$0$
with respect to the orbital state.
We observe a sharp ferro-orbital peak of $T(q)$ at $q$=$0$,
while $T(0)$ is found to decrease as the time evolves,
indicating that the orbital state changes
due to the dynamical charge excitation.
This is because the pair-hopping process contributes to
the doublon motion.
Indeed, once a doublon is created in the $3z^2$$-$$r^2$ orbital,
it can transfer to the $x^2$$-$$y^2$ orbital within a site
through the pair-hopping process.
Thus, the orbital configuration is dynamically deformed.

\begin{figure}[b]
\includegraphics*[width=\linewidth]{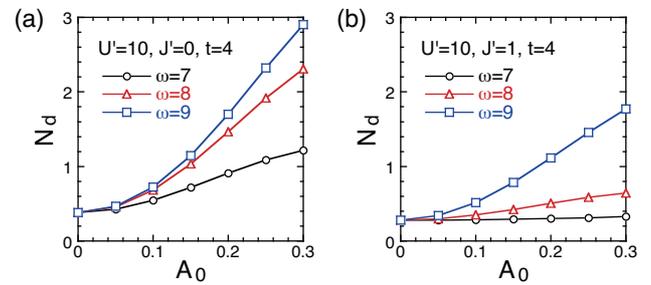}
\caption{%
The $A_{0}$ dependence of the doublon number $N_{\rm d}$ at the early time $t$=$4$
for several values of $\omega$.
(a) $N$=$16$, $U'$=$10$, and $J'$=$0$,
and (b) $N$=$16$, $U'$=$10$, and $J'$=$1$.
}
\label{fig4}
\end{figure}

Here, we study
how the dynamical charge excitation is affected
by the amplitude of the vector potential $A_{0}$
in the weak perturbation regime.
In Fig.~4,
we show the $A_{0}$ dependence of $N_{\rm d}$
at the early time $t$=$4$
for several values of $\omega$ at $J'$=$0$ and $1$.
At a first glance,
we can see that
$N_{\rm d}$ monotonously increases
with increasing $A_{0}$
regardless of the values of $\omega$ and $J'$.
This observation can be naturally understood,
since the dynamical charge excitation should become significant
when the strong electric field is applied.
Note that $N_{\rm d}$ increases as $\omega$ increases
at fixed $J'$ and $A_{0}$,
as we have seen in Figs.~2 and 4.

\section{Summary and discussion}

We have examined the nonequilibrium charge-spin-orbital dynamics
under an oscillating electric field,
by analyzing the one-dimensional $e_{\rm g}$-orbital Hubbard model.
We have clearly observed the dynamical charge excitation
accompanied by the creation of holons and doublons.
As the time evolves,
the doublons propagate through the pair-hopping process,
leading to the dynamical deformation of the orbital configuration.

In the previous work
\cite{Onishi2010},
we have investigated the time evolution of a holon-doublon wavepacket
in the $e_{\rm g}$-orbital Hubbard model without the electric field.
We have found that the pair-hopping process contributes to the doublon propagation,
and thus the orbital structure is dynamically deformed.
The present results are fully consistent with this previous observation.

Regarding the adopted model Hamiltonian,
we note that
the hopping amplitude is spatially anisotropic
such that the $3z^2$$-$$r^2$ orbital is itinerant,
while the $x^2$$-$$y^2$ orbital is localized.
Here, we evaluate the hopping amplitude
from the overlap integrals of $e_{\rm g}$-orbital wavefunctions
through the $\sigma$-bond $(dd\sigma)$
\cite{Slater1954}.
When we also include higher-order $(dd\pi)$ and $(dd\delta)$,
$t_{\alpha\alpha}$=$(dd\sigma)$,
$t_{\beta\beta}$=$(dd\delta)$,
and
$t_{\alpha\beta}$=$t_{\beta\alpha}$=$0$.
That is,
the $x^2$$-$$y^2$ orbital turns to be itinerant with small hopping amplitude.
In such a situation,
electrons tend to occupy
not only the $3z^2$$-$$r^2$ orbital but also the $x^2$$-$$y^2$ orbital
in the ground state.
After the electric field is turned on,
doublons are created and propagated in both orbitals,
and thus the electron hopping process would contribute to
the deformation of the orbital structure.


\begin{acknowledgement}
Part of the computations were done
on the supercomputer at the Japan Atomic Energy Agency.
This work was supported by a Grant-in-Aid for
Scientific Research of the Ministry of Education,
Culture, Sports, Science, and Technology of Japan.
\end{acknowledgement}


%

\begin{thebibliography}{[1]}

\bibitem{Tokura2006}%
Y. Tokura, J. Phys. Soc. Jpn. \textbf{75}, 011001 (2006).

\bibitem{Miyano1997}%
K. Miyano, T. Tanaka, Y. Tomioka, and Y. Tokura,
Phys. Rev. Lett. \textbf{78}, 4257 (1997).

\bibitem{Fiebig1998}%
M. Fiebig, K. Miyano, Y. Tomioka, and Y. Tokura,
Science \textbf{280}, 1925 (1998).

\bibitem{Nasu2004}%
K. Nasu,
Rep. Prog. Phys. \textbf{67}, 1607 (2004).

\bibitem{Takahashi2008}%
A. Takahashi, H. Itoh, and M. Aihara,
Phys. Rev. B \textbf{77}, 205105 (2008).

\bibitem{Koshibae2009}%
W. Koshibae, N. Furukawa, and N. Nagaosa,
Phys. Rev. Lett. \textbf{103}, 266402 (2009).

\bibitem{Onishi2010}%
H. Onishi,
J. Phys.: Conf. Ser. \textbf{200}, 012152 (2010).

\bibitem{Maeshima2010}%
N. Maeshima, K. Hino, and K. Yonemitsu,
Phys. Rev. B \textbf{82}, 161105 (2010).

\bibitem{Matsueda2012}%
H. Matsueda, S. Sota, T. Tohyama, and S. Maekawa,
J. Phys. Soc. Jpn. \textbf{81}, 013701 (2012).

\bibitem{Kanamori2012}%
Y. Kanamori, J. Ohara, and S. Ishihara,
Phys. Rev. B \textbf{86}, 045137 (2012).

\bibitem{Slater1954}
J.C. Slater and G.F. Koster,
Phys. Rev. \textbf{94}, 1498 (1954).

\bibitem{Note1}%
A typical value of the electron hopping in
$3d$ transition-metal oxides such as manganites is
$t_{\alpha\alpha}$$\sim$$0.4$~eV,
and then the time $t$=$10$ corresponds to $16$~fs.

\bibitem{White1992}%
S.R. White,
Phys. Rev. Lett. \textbf{93}, 2863 (1992).

\bibitem{Daley2004}%
A.J. Daley, G. Kollath, U. Schollw\"ock, and G. Vidal,
J. Stat. Mech.: Theory Exp., P04005 (2004).

\bibitem{White2004}%
S.R. White and A.E. Feiguin,
Phys. Rev. Lett. \textbf{93}, 076401 (2004).

\bibitem{Schollwock2005a}
U. Schollw\"ock,
Rev. Mod. Phys. \textbf{77}, 259 (2005).

\bibitem{Schollwock2005b}
U. Schollw\"ock,
J. Phys. Soc. Jpn. \textbf{74}, Suppl. 246 (2005).

\end{thebibliography}
%

\end{document}